\documentstyle{elsart}

\begin{document}

\begin{frontmatter}

\title{QCDF90: Lattice QCD with Fortran 90\thanksref{label}}

\thanks[label]{This research is supported in part under DOE grant
               DE-FG02-91ER40676.
               We are grateful to the Center of Computational Science 
               and the Office of Information Technology for support 
               and access to the Boston University supercomputer 
               facility.}

\author[A]{ I.~DASGUPTA }
\author[A]{ A.R.~LEVI }
\author[A]{ V.~LUBICZ }
\author[A]{ and C.~REBBI }

\address[A]{\it Department of Physics, Boston University,
            590 Commonwealth Avenue, Boston, MA 02215, USA}

\begin{abstract}
We have used Fortran 90 to implement lattice QCD.
We have designed a set of machine independent modules
that define fields (gauge, fermions, scalars, etc...)
and overloaded operators for all possible operations 
between fields, matrices and numbers. 
With these modules it is very simple to write high-level
efficient programs for QCD simulations. 
To increase performances our modules also implements
assignments that do not require temporaries, 
and a machine independent precision definition. 
We have also created a useful compression procedure for 
storing the lattice configurations, and 
a parallel implementation of the random generators.
We have widely tested our program and modules on 
several parallel and single processor supercomputers 
obtaining excellent performances.
\end{abstract}

\end{frontmatter}

\section{Introduction}

Simulating Quantum Chromo Dynamics (QCD) on the lattice is a 
challenging task for
the computers of today. The simulations involve a huge
number of variables and are so demanding on the computer resources that
QCD simulation 
is already regarded as a benchmark test for the efficiency and
performance of the modern supercomputers.
With the twofold goal of facilitating the development of algorithms
and applications for lattice QCD,
and of maintaining good code performance, we have taken advantage of
the possibilities offered by Fortran 90 to write a set of modules for
a high-level, yet efficient implementation of QCD simulations.
Fortran 90 offers the possibility to
define both types and overloaded operators.
These two key features make Fortran 90 particularly suitable
for writing QCD programs at a very high level where the algorithms are
transparent and compact. On the other hand, by maximizing efficiency of
the code in the definition of the overloaded 
operators, uniform and automatic efficiency is ensured in the entire
simulation with little effort required from the high level programmer.

Our end product is a package, ``QCDF90'',
which is fully described in a long documentation [1],
where we provide all the information needed to use our package.
In the following we will highlight the main 
characteristics of QCDF90.

\section{Precision }

\noindent
To render, from the beginning, the precision definitions 
machine independent, the module ``precision''
defines two  $\tt kind $ parameters, $\tt REAL8 $ and $\tt LONG $.
These parameters store the $\tt kind $ of an 8-byte 
floating point variable and of an 8-byte integer variable.

\section{Lattice geometry }

\noindent
The set of all lattice sites can be subdivided into ``even'' 
and ``odd'' sites according to whether the sum of the integer 
valued coordinates
$\tt x+y+z+t$ is even or odd (checkerboard subdivision). 
This division is actually quite useful in lattice algorithms, 
especially in the context of parallel implementations.
In QCDF90 we implement the same checkerboarded separation of 
the lattice in two sublattices. Correspondingly all field variables 
are divided into even and odd variables. 
We encode this important division 
in the very first component of the type definitions for the fields. 
This component is an 
integer variable called $\tt parity$ which takes the 
values $\tt 0$ and $\tt 1$ for variables defined over even and odd 
sites respectively.

For the gauge variables and for the variables associated with
the generators it is convenient
to define a separate type for all four directions. 
This is realized with a second integer component 
variable $\tt dir$ which takes values from $\tt 1$ 
to $\tt 4$ for variables defined in the corresponding direction.
 
In a vectorized or superscalar architecture pipelined instructions 
and longer arrays give origin to better performance. 
Therefore the lattice is most efficiently indexed by 
a single lattice volume index ranging
$\tt xyzt$ from $0$ to $NX*NY*NZ*NT/2-1$ for each sublattices
(where $Ni$ is the lattice size in direction $i$).

\section{Field definitions }

Here is a summary of all the types we have defined in the modules:
$\tt gauge\_field$ (a $3*3$ complex matrix in a given direction);
$\tt fermi\_field$ (a $3$ component complex vector times $4$ 
spinor indices);
$\tt complex\_field$ (a complex scalar);
$\tt real\_field$ (a real scalar);
$\tt generator\_field$ ($8$ real variables in a given direction, 
for the $SU(3)$ generators).
All the above field are defined on an even or odd sublattice. 

In addition we define the type
$\tt full\_gauge\_field$ (a collection of $8$ gauge\_fields) to store
the entire gauge field configuration (including both parities and all
directions).
Moreover we defined
the type $\tt matrix $ as a $3*3$ complex matrix.

\section{Overloaded operators}

\noindent
To complete the high-level structure of the code one needs to define
high level operators between the types described above.
We define overloaded operators for all possible operations 
between fields, matrices, complex and real numbers. 
The overloaded operator set includes multiplication, 
multiplication with the adjoint, division,
addition, subtraction, lambda matrices algebraic manipulations,
gamma matrices algebraic manipulations, adjoin-ing, conjugation,
real and complex traces, exponentiation, square root, contraction, 
etc...

For example, if $\tt g_i$ are gauge fields, the use of 
overloaded operators allows instructions to be as
simple as:

$\tt g_1=g_2*g_3$,

\noindent
specifically, such an instruction means that, at each site of
the hipercubical lattice, the  $3 \times 3$ complex
matrix multiplication between $\tt g_2$ and $\tt g_3$
is performed.

We have maximized code efficiency in the definitions of the overloaded
operators. In every case, the overloaded 
operation is implemented with the minimum
number of elementary arithmetic operations (see the section on
Assignments).

Further operators have been overloaded to
perform special operations involving the $SU(3)$ 
$\tt generator\_fields$ and $\tt gauge\_fields$.  
In particular it is very important to have an efficient algorithm 
for the matrix exponentiation, since this operation can be a time
consuming component of several QCD calculations. We use a new algorithm
that takes advantage of the properties of the $3*3$
Hermitian traceless matrices to perform the exponentiation
with a minimal number of arithmetic operations.

\section{Shifts and parallel transport}

\noindent
Shifts are also implemented as overloaded operators. 
For each field type C-shift implements an ordinary shift 
of the field with respect to the Cartesian geometry of the lattice.  

Moreover, because gauge theories are characterized by the
property of local gauge invariance, we find it very useful to
directly define a U-shift operator that consists of the 
shift with the appropriate parallel transport factor.
In fact in a gauge theory the parallel transport (U-shift)
 is the relevant shift operator
and is the natural building block for programming.

The efficient manipulation of the Dirac operator
is the most critical issue in QCD. 
For the Fermi fields we found it useful to
define other shift operators which incorporate fundamental 
features of the Dirac operator.
Firstly, one needs a shift operator that combines the naive shift with 
the parallel transport and the appropriate gamma matrix
manipulation. This is implemented by the overloaded operators W-shift. 
Acting on a Fermi field $\tt f_1$, W-shift in the
direction $\mu$ produces a
Fermi field $\tt f_2$, given by 
\begin{equation}
f_{2,{\bf x}}= (1-\gamma_{\mu})U^{\mu}_{{\bf x}} 
f_{1,{\bf x}+\hat\mu}
\end{equation}
for positive $\mu$, and
\begin{equation}
f_{2,{\bf x}}= (1+\gamma_{\mu})
U^{\dagger \mu}_{{\bf x}-\hat\mu} f_{1,{\bf x}-\hat\mu}
\end{equation}
for negative  $\mu$. 
In these equations $U^{\mu}_{{\bf x}}$ and 
$U^{\dagger \mu}_{{\bf x}-\hat\mu}$ are the gauge link
variable necessary to implement the proper parallel transport.
Combining shift with gamma matrix manipulation in the 
operator W-shift pays off by conserving computer resources,
in fact, it follows from the properties of the gamma
matrices that only one half of the spin components of
the field $f_1$ undergo the transport at a time. 
The W-shift is the minimal specialized 
operator that achieves this transport.

The complete lattice Dirac-Wilson operator has been also overloaded, 
as well as the adjoint operator
Xdirac$=\gamma_5$ Dirac $\gamma_5$.

\section{Assignments}

\noindent
The use of overloaded operators may imply the creation of more 
temporaries and, consequently, more motion of data than a 
straightforward implementation of operations among arrays.  
Consider for example the following operation among variables 
of type fermi\_field: $\tt f1=f1+f2+f3 $.
As far as we know, Fortran 90 does not specify how the 
variables should be passed in function calls.
As a consequence, the above instruction may require as many 
as four temporaries depending on the operating system. (An operating
system that implements overloaded operations via function calls would
first add $\tt f1$ and $\tt f2$, placing the result in a temporary $\tt
t1$ whose address would then be passed to the calling program. The
compiler would then copy $\tt t1$ into a temporary $\tt t2$, add $\tt
f3$ to $\tt t2$ and place the result in $\tt t1$. Finally $\tt t1$ would
be copied into $\tt f1$.)
The procedure could be drastically simplified and made system
independent through the use of
an overloaded assignment $\tt +=$.  
The above instruction could be written $\tt f1~ += f2+f3 $
which the compiler would implement by issuing first a call to 
a function that adds $\tt f2$ and $\tt f3$ returning the result 
in $\tt t1$.  The addresses of $\tt f1$ and $\tt t1$ would then 
be passed to a subroutine that implements the operation 
$\tt f1=f1+t1$ among the components of the data types.  
The required number of copies to memory would be only two, 
instead of four and it does not depend on the operating system. 

In order to allow for these possible gains in efficiency, we 
have defined a large set of overloaded assignments.
Since Fortran 90 permits only the use of the $\tt =$ symbol 
for the assignment, we have defined two global variables: a character 
variable $\tt assign\_type$ and an integer variable 
$\tt assign\_spec$. The latter is introduced to accommodate assignments
of a more elaborate nature. 
The default values of these variables are ``$\tt =$'' and 
``$\tt 0$''.  
Overloaded assignments are obtained by setting 
the $\tt assign\_type$ (and, if necessary, the
$\tt assign\_spec$) to the 
appropriate value immediately before the assignment. 
Our example then becomes $\tt assign\_type='+'; f1=f2+f3 $. The use of
the variable $\tt assign\_spec$ can be illustrated with another example. 
Consider the U-shift operation that implements shifts on gauge
fields. There are in fact four U-shifts, corresponding to the four
directions for the shift. To shift the gauge field $\tt b$ in the
direction $\tt n$ and copy the result in the gauge field $\tt a$ we will
use: 

$\tt assign\_type='u'; assign\_spec= n; a=b $

Note that the $\tt assign\_type $ and the $\tt assign\_spec$ are
automatically reset to their default values
after each use. 
Therefore every call to an overload assignment operation must be
preceded by an assignment statement.
This ensure protection against misuse of the assignment.

Assignments can be defined between variables of different types. 
We have collected these assignments in 
the module ``assign\_mixed''. Often these assignments have very useful
but not necessarily obvious definitions. 
For example if $\tt c_1$ is a complex\_field and 
$\tt complex$ is a complex variable, the instruction
$\tt complex=c_1$, is interpreted as setting the variable 
$\tt complex$ to the sum over all the lattice of the 
components of $\tt c_1$.

References [1] provides a detailed explanation of all 
the assignment instructions.

\section{Conditionals}

\noindent
The module ``conditional''
defines six overloaded relational operators, $\tt >$, $\tt >=$,
$\tt < $, $\tt <= $, $\tt == $, $\tt /=$, and the
$\tt .Xor.$ operator.

The relational operators perform 
two tasks: first they return a logical variable 
(true if $\tt parity$ and $\tt dir$ components of the operands are
the same, false if they are undefined or not the same),
second, and more important, they set 
the global variable $\tt context$ to 
$\tt .TRUE.$ at all sites where the relation is satisfied
and to $\tt .FALSE.$ at all other sites. 

The operator $\tt .Xor.$ admits as operands a pair of fields
of the same type and returns a field, also of the same type,
having as elements the corresponding elements of the first operand 
at the sites where the global variable $\tt context$ is $\tt .TRUE.$ and 
the elements of the second operand at the sites where 
$\tt context$ is $\tt .FALSE.$. 

This can be used to select elements out of two fields according 
to some local condition, an operation which lies at the 
foundation of stochastic simulation techniques.

\section{Random numbers}

We have implemented a parallelizable version of the unix pseudorandom
number generator erand48, which also provides added functionality.
Erand48 is a congruential pseudorandom number generator based 
on the iterative formula
\begin{equation}
s_{i+1}=a_1*s_i+b_1 \quad {\rm mod} \; 2^{48} \ ,
\label{erand}
\end{equation}
where $a_1=\tt 0x5DEECE66D$, $b_1=\tt 0xB$, $s_i$ 
and $s_{i+1}$ are integers of at least 48 bits of precision.  
The ``seeds'' $s_i$ are converted
to real pseudorandom numbers $r_i$ with uniform distribution 
between $0$ and $1$ by $r_i=2^{-48}\, s_i$.

As presented above, the algorithm is intrinsically serial. 
However it follows from Eq.~(\ref{erand}) that the 
$\rm N^{th}$ iterate $s_{i+N}$ is still of the form 
\begin{equation}
s_{i+N}=a_N*s_i+b_N \; {\rm mod} \; 2^{48}
\end{equation}
with integers
$a_N$ and $b_N$ which are uniquely determined by $a_1$, $b_1$.
The module takes advantage of this fact and of the definitions of
a global variable $\tt seeds$ to generate 
pseudorandom numbers in a parallelizable fashion.  
The module ``random\_numbers'' overloads operators that generate
Gaussian or uniformly distributed fields and do all the 
necessary seed manipulations.

\section{Write and read configurations}

\noindent
To store and retrieve an entire SU(3) gauge field 
configuration, we have developed a portable, compressed ASCII format.  
Only the first two columns of the gauge field matrices are stored, 
thanks to unitarity and unimodularity.
Our subroutines takes advantage of the fact
that all of the elements of the gauge field matrices have magnitude
smaller or equal to 1 to re-express their real and imaginary parts 
as 48bit integers.  These integers are then written in base 64,
with the digits being given by the ASCII collating sequence
starting from the character ``0'' (to avoid unwanted ASCII characters).
Thus, an entire gauge field matrix is represented 
by 96 ASCII characters, without loss of numerical information.  

To ensure the recover of the data,
a detailed description of the compressing  procedure is written, 
as a header, at the beginning of each stored configuration file itself.

\section{Performances}

The code has been designed to run in any computer.
It has been tested 
on a SGI PowerChallengeArray 
with 90 MHz processor nodes, using IRIX 6.1 and IRIX 6.2
and Fortran 90,
with a single processor -O3 optimization flags or 
with the flags -O3 -pfa -mp to implement multiprocessing;
on a Silicon Graphics Indigo using the IRIX 6.1 Fortran 90;
and on the IBM R6000 58H model 7013 at 55 MHz, 
running AIX
with the xlf90 IBM compiler using the -O3 optimization flags.

Memory required to execute varies according to the applications.
Scales proportionally to the lattice volume $NX*NY*NZ*NT$.
On a $16^4$ lattice, the example codes included in QCDF90,
quenched.f90 and propagator.f90 use
approximately 110 Mbytes and 140 Mbytes respectively.

The run of the example programs
quenched.f90 and propagator.f90 
take approximately $45$ microsec to update an SU(3) link, and
$8$ microsec to calculate a plaquette, 
and $20$ microsec for a CG step per link, using a $16^4$ lattice,
on an SGI Power-Challenge per node.

\end{document}